\documentclass[12pt,preprint]{aastex}

\makeatother

\newcommand{\s}{{\it Spitzer}}
\newcommand{\h}{{\it Herschel}}
\newcommand{\msol}{M$_{\odot}$}
\newcommand{\md}{$M_{\rm dust}$}

\newcommand{\lsol}{L$_{\odot}$}
\newcommand{\lir}{$L_{\rm IR}$}
\newcommand{\lco}{$L^{\prime}_{\rm CO}$}

\newcommand{\td}{$T_{\rm d}$}
\newcommand{\aco}{$\alpha_{\rm CO}$}
\newcommand{\gd}{$M_{\rm gas}/M_{\rm d}$}
\newcommand{\mgas}{$M_{\rm gas}$}
\lefthead{Magdis et al..~}

\slugcomment{to appear in ApJ}
\shortauthors{Magdis et al.}
\begin{document}
 \title{GOODS-\h~: Gas-to-dust mass ratios and CO-to-H$_{2}$ conversion factors in normal and starbursting galaxies at high-z}
 \author{Georgios E. Magdis $\!$\altaffilmark{1,2},
 	     E. Daddi $\!$\altaffilmark{2},
		 D. Elbaz $\!$\altaffilmark{2},        
		 M. Sargent $\!$\altaffilmark{2},
         M. Dickinson $\!$\altaffilmark{3},
         H. Dannerbauer $\!$\altaffilmark{2},
         H.~Aussel $\!$\altaffilmark{2},
		 F.  Walter $\!$\altaffilmark{4},
		 H.S. Hwang $\!$\altaffilmark{2},
		V.~Charmandaris $\!$\altaffilmark{5,6,7},	     
	      J. Hodge $\!$\altaffilmark{4},
         D. Riechers $\!$\altaffilmark{8},
         D. Rigopoulou $\!$\altaffilmark{1},        
         C. Carilli $\!$\altaffilmark{9},
		 M.~Pannella $\!$\altaffilmark{2},        
         J.~Mullaney $\!$\altaffilmark{2},
         R.~Leiton $\!$\altaffilmark{2},
         D.~Scott $\!$\altaffilmark{10}
}          
\altaffiltext{1}{Department of Physics, University of Oxford, Keble Road, Oxford OX1 3RH}
\altaffiltext{2}{CEA, Laboratoire AIM, Irfu/SAp, F-91191 Gif-sur-Yvette, France}
\altaffiltext{3}{NOAO, 950 N. Cherry Avenue, Tucson, AZ 85719, USA}
\altaffiltext{4}{Max-Planck-Institut f\"ur Astronomie, K\"onigstuhl 17, D-69117 Heidelberg, Germany}
\altaffiltext{5}{Department of Physics \& ICTP, University of Crete, GR-71003, Heraklion, Greece}
\altaffiltext{6}{IESL/Foundation for Research \& Technology-Hellas, GR-71110, Heraklion, Greece}
\altaffiltext{7}{Chercheur Associ\'e, Observatoire de Paris, F-75014,  Paris, France}
\altaffiltext{8}{Astronomy Department, California Institute of Technology, MC 249-17, 1200 East California Boulevard, Pasadena, CA
91125, USA}
\altaffiltext{9}{National Radio Astronomy Observatory, PO Box O, Socorro,NM 87801, USA}
\altaffiltext{10}{Department of Physics and Astronomy, University of British Columbia, Vancouver, BC V6T 1Z1, Canada}

\begin{abstract}
We explore the gas-to-dust mass ratio (\gd) and the CO luminosity-to-\mgas ~conversion factor (\aco) of two well
studied galaxies in the GOODS-N field, that are expected to have different star forming modes, the starburst GN20 at $z=4.05$
and the normal star-forming galaxy BzK-21000 at $z=1.52$. Detailed sampling is available for their Rayleigh-Jeans
emission via ground based mm interferometry ($1.1-6.6\,$mm) along with \h ~PACS and SPIRE data that probe the peak
of their infrared emission. Using the physically motivated Draine \& Li (2007) models, as well as a modified
black body function, we measure the dust mass (\md) of the sources and find $(2.0^{+0.7}_{-0.6} \times10^{9})$ \msol
~for GN20 and $(8.6^{+0.6}_{-0.9} \times 10^{8})$ \msol ~for BzK-21000. The addition of mm data reduces the
uncertainties of the derived \md ~by a factor of $\sim2$, allowing the use of the local  \gd ~vs metallicity relation
to place  constraints on the \aco ~values of the two sources. For GN20 we derive a conversion factor of \aco ~$< $ 1.0 \msol ~pc$^{-2}$(K km s$^{-1}$)$^{-1}$, consistent with that of local ULIRGs, while for BzK-21000 we find a considerably higher value, \aco ~$\sim4.0$ \msol pc$^{-2}$(K km s$^{-1}$)$^{-1}$, in agreement with an independent kinematic derivation reported previously. The implied star
formation efficiency is $\sim25$ L$_\odot$/M$_\odot$ for BzK-21000, a factor of $\sim5-10$ lower than that of GN20.
The findings for these two sources support the existence of different disk-like and starburst
star-formation modes in distant galaxies, although a larger sample is required to draw statistically robust results.
\end{abstract}

\section{Introduction}

 The determination of  the conversion factor from CO luminosities to the molecular gas mass (\mgas) of a galaxy, (\aco\footnote{The units of \aco, \msol pc$^{-2}$ (K km s$^{-1}$)$^{-1}$, are omitted from the text for clarity.}~= \mgas ~/ \lco) 
 remains an open issue as there is evidence that it varies considerably as a function of metallicity and intensity of the radiation field.  
 Downes \& Solomon (1998) showed that \aco ~is a factor of $\sim 6$ smaller for local ultra-luminous infrared galaxies (ULIRGs, $L_{\rm IR} > 10^{12} L_\odot$) than for local spiral galaxies. A similar picture seems to emerge at high redshift, with a fraction of submillimeter galaxies (SMGs) playing the role of
local starbursts, having lower \aco ~values and exhibiting enhanced star formation efficiencies (defined as \lir/\mgas), when compared to normal\footnote{Throughout this paper, we use the term ``normal''  to refer to star-forming galaxies that fall within the so-called ``main sequence'' relation between their star formation rates and stellar masses (e.g., Brinchmann et al. 2004; Noeske et al. 2007; Elbaz et al. (2007,2011); Magdis et al. 2010a).   Instead, ``starbursts'' are galaxies with substantially elevated SFRs for their stellar masses.} star-forming galaxies selected by their rest-frame UV or optical light (Tacconi et al. 2008, Daddi et al. 2010, Genzel et al. 2010,
Narayanan et al. 2011). Daddi et al. (2010a,b) used a kinematic analysis of star-forming disk galaxies at $z \approx 1.5$ to argue that they have    CO conversion factors $\alpha_{\mathrm CO} = 3.6 \pm 0.8$, like that in the Milky Way. Instead, Tacconi et al. (2008) and Carilli et al.\ (2010),  placed an upper limit of $\sim0.8$ on the \aco ~value of high$-z$ SMGs. Together with a variety of other evidence, these results support the existence of two distinct star formation regimes: a long-lasting mode for normal-disks galaxies and a more rapid mode for local starbursts and  SMGs (Daddi et al. 2008, 2010; Genzel et al. 2010). Nevertheless, excitation biases
introduced by the use of different molecular lines, along with substantial
uncertainties on the \aco ~values, raise potential concerns about the role of SMGs in this picture (e.g., Ivison et
al. 2011).

Several studies in the local Universe have tried to tackle this question by measuring the 
total dust mass of a galaxy (\md) and assuming that it is proportional to \mgas ~(e.g., Leroy et al. 2011).
However, determining \md ~is a complex task. In order to break degeneracies inherent in current models, a
proper characterization of the peak of the spectral energy distribution (SED) of the source as well as of the
Rayleigh-Jeans emission tail is required.  With the wide wavelength coverage (70 to 500$\,\mu$m ) of the
Herschel Space Observatory (\h, Pilbratt et al. 2010), we can now directly observe the peak of the IR
emission of high$-z$ galaxies. Together with ground based mm observations that probe the Rayleigh-Jeans emission, this
allows us  to properly quantify \md,  and explore possible differences between the shape of the SED, the \gd ~ratios and the \aco ~values of normal and starburst galaxies at high redshift. 

As a test case, we combine \h ~PACS and SPIRE data, from the GOODS-\h ~program with (sub)mm observations for two of the best-studied high$-z$ galaxies in the Great Observatories Origins Deep Survey North field (GOODS-N), the SMG GN20 at $z=4.05$, and the normal star-forming galaxy, BzK-21000, at $z=1.521$. The uniqueness of these sources relies on the detailed sampling of their Rayleigh-Jeans emission via ground based mm interferometry ($1.1-6.6\,$mm). Our aim is to derive \gd ~ratio estimates, investigate the slope of their SED in the Rayleigh-Jeans tail and put constraints on the \aco ~values. Throughout this paper we assume $\Omega_{\rm m}=0.3$, H$_{0}=71$ km sec$^{-1}$ Mpc$^{-1}$, $\Omega_{\Lambda}=0.7$ and Chabrier IMF.


\section{Sample and observations}
We use  deep 100- and $160\,\mu$m PACS and 250-, 350-, $500\,\mu$m SPIRE observations of GOODS-N from the
GOODS-H program. Details about the observations are given in Elbaz et al. (2011). Herschel 
fluxes are derived from PSF fitting using {\em galfit} (Peng et al. 2002). A very extensive set of priors
is used for 100-, 160- and $250\,\mu$m, including all galaxies detected in the ultra-deep \s ~MIPS $24\,\mu$m imaging, 
which effectively allow to obtain robust flux estimates for relatively isolated sources, even beyond formal confusion 
limits at 250$\,\mu$m. For 350- and 500$\,\mu$m this approach does not allow accurate measurements due to the increasingly
large PSFs. Hence we use a reduced set of priors based on VLA radio detections, resulting in flux uncertainties consistent with the
 confusion noise at these wavelengths. We note that GN20, the radio priors include also the nearby GN20.2 objects (both``a'' and ``b" components; Daddi et al. 2009). A detailed description of the flux
measurements and Monte Carlo derivations of the uncertainties will be presented elsewhere (Daddi et al. in
preparation). 

Originally detected at 850$\,\mu$m by Pope et al. (2006), GN20 is one of the best studied SMGs to date,
the most luminous and also one of the most distant ($z=4.055$; Daddi et al. 2009a) in the GOODS-N field. 
It is detected in all \h ~bands apart from 100$\,\mu$m. 
Carilli et al. (2010), reported the detection of the CO[1$-$0] and CO[2$-$1] lines with the Very Large Array (VLA), and CO[6$-$5] and CO[$5-$4] lines with the Plateau de Bure Interferometer (PdBI) and the Combined Array for Research in Millimeter Astronomy (CARMA) respectively. The source is detected in the Aztec 1.1$\,$mm map and continuum emission is also measured at 2.2-, 3.3-, and 6.6$\,$mm (Carilli et al. 2011) and at 1.4$\,$GHz with the VLA (Morrison et al. 2009). GN20 is identified in all IRAC bands and at 24$\,\mu$m while it appears as a $B$-band dropout in the ACS-HST image. The stellar mass of the source is $2.3 \times 10^{11}$ \msol ~(Daddi et al. 2009a). A compilation of the photometric data is given in Table 1.

BzK-21000 is a near-IR selected star-forming galaxy, with a spectroscopic redshift, $z=1.521$
(Daddi et al. 2008, 2010a). The source has secure detections in both PACS and in the first two SPIRE bands, while at
500$\,\mu$m it is only marginally detected. In addition to IRAC and MIPS $24\,\mu$m data, the source is seen in the
16$\,\mu$m IRS peak-up image (Teplitz et al. 2011). With follow-up VLA and PdBI observations,  Dannerbauer et al.
(2009), Aravena et al. (2010) and Daddi et al. (2008), have reported the detection of CO[3$-$2], CO[1$-$0] and  CO[2$-$1]
emission lines respectively. Continuum detections and upper limits are also obtained at 1.1-, 2.2- and 3.3$\,$mm (Daddi et
al. 2010, Dannerbauer et al. 2009). The stellar mass is $7.8 \times 10^{10}$ \msol ~(Daddi et al. 2010a). The UV rest-frame morphology of this galaxy, the double-peaked CO profile, the large spatial extent of the CO reservoir, and the low gas excitation all provide strong evidence that this galaxy is a large, clumpy, rotating disk (Daddi et al. 2010a).

\section{Estimating  total dust masses}
We employ two methods to derive the dust mass of the galaxies:  the physically motivated dust models of Draine \& Li (2007) (DL07 hereafter), and a more simplistic, but widely used, modified black body model (MBB).

The DL07 models describe the interstellar dust as a mixture of carbonaceous grains and amorphous silicate grains. The properties of these grains are 
parametrized by the PAH index, $q_{\rm PAH}$, defined as the fraction of the dust mass in the form of PAH grains. The majority of the 
dust is supposed to be located in the diffuse ISM, heated by a radiation field with a constant intensity $U_{\rm min}$. A smaller 
fraction $\gamma$ of the dust is exposed to starlight with intensities ranging from  $U_{\rm min}$ to $U_{\rm max}$, representing 
the dust enclosed in photo-dissociation regions (PDRs). Following the prescription of DL07, we fit the rest-frame mid-IR to mm data points and search for the best-fit model by minimizing the reduced $\chi^{2}$. The total dust mass is derived from the best-fit model and its uncertainty is estimated by the distribution of $M_{\rm d}$ values that correspond to models with $\chi^{2}$ $\le$ $\chi^{2}_{min}+1$ (Avni 1976). Although a different grain size distribution would result in different \md, we choose to adopt the one prescribed by the DL07 models as they can successfully reproduce the IR and submillimeter emission for a sample of SINGS galaxies (including both normal and starburst galaxies).

We also fit the SEDs of our sources with the standard form of a single temperature (\td) modified black body, leaving the effective emissivity  ($\beta_{\rm eff}$) as a free parameter along with the dust temperature (\td). For the fit we only consider data points with $\lambda_{rest}>40\,\mu$m, to avoid emission from very small grains and from the best fit model we can then estimate the \md ~from the relation:
\begin{equation}
M_d = \frac{S_\nu D^2_L}{(1+z)\kappa_{rest} B_\nu (\lambda_{rest} , T_d)}
\end{equation}
where $S_{\nu}$ is the observed flux density, 
$D_{\rm L}$ is the luminosity distance and $\kappa_{\rm rest}$ = $\kappa_{\rm 0} (\lambda_{\rm 0}/\lambda_{\rm rest})^{\beta}$  is the rest frame dust mass absorption coefficient at the observed wavelength (Li \& Draine 2001). The uncertainty in \md ~is 
obtained as for the DL07 models. The photometric data, along with the best fitting DL07 and MBB models are shown in Fig. 1, while in Table 2 we summarize the derived parameters.

The two methods return \md ~estimates that are in broad agreement. In particular, for GN20 we derive  \md ~=
2.0$^{+0.7}_{-0.6}$ $\times$ 10$^{9}$ \msol (DL07) and 1.5$^{+0.4}_{-0.5}$ $\times$ 10$^{9}$ \msol (MBB) while for
BzK-21000 the corresponding values are \md ~= 8.6$^{+0.6}_{-0.9}$ $\times$ 10$^{8}$ and 7.6$^{+1.2}_{-1.3}$ $\times$ 10$^{8}$ \msol. Best fit MBB models also indicate \td ~= 33K and 34K for GN20 and BzK-21000 respectively and $\beta_{\rm eff}=2.1\pm0.2$ and $1.4\pm0.2$.   
The dependence of \md ~on the other two free parameters of the MBB models\footnote{These values are derived under the assumption of optical thinness. If we drop this assumption the corresponding values are \td~=33.8 and 46.3
K for BzK2100 and GN20.}, i.e. \td ~and $\beta_{\rm eff}$, along with the 68\% and 99\% confidence intervals, are shown in Fig. 2. 

To evaluate the significance of adding mm data in the derivation of \md,
we repeat the fitting procedure, this time excluding any data at wavelengths longer than 850$\,\mu$m. For the DL07 models, the best fit \md ~are unaffected, but the uncertainties increase by a factor of $\sim2$. Similar results are presented by several studies in the local universe (e.g. Draine et al. 2007,  Galametz et al. 2011),  where they find that in the absence of rest-frame (sub)mm data, the derived \md ~estimates and highly uncertain.

\section{Discussion}

We derive a total IR luminosity of \lir ~= $(1.9 \pm 0.4) \times 10^{13}$ \lsol ~for GN20 and
 \lir ~= $(2.1 \pm 0.3) \times 10 ^{12}$ \lsol ~for BzK-21000, that correspond to SFRs of $\sim2000$ and $\sim210$ \msol
  ~yr$^{-1}$ and specific star formation rates (sSFR) of $\sim8.6$ and $\sim2.6$ Gyr$^{-1}$. Several
   studies have shown that star-forming galaxies at any redshift follow a tight SFR$-$M$_\ast$ relation, with outliers being
    starburst galaxies (e.g., Brinchmann et al. 2004,
Elbaz et al. 2007,  Daddi et al. 2007, 2009, Magdis et al. 2010a). Our results confirm the sSFR of BzK-21000 is similar to that of main sequence galaxies defined in the SFR$-$M$_\ast$ space at this redshift, while GN20 is located in the starburst regime.

As illustrated in Fig. 1, it appears that GN20 has a steeper slope in the Rayleigh-Jeans tail when compared to that of BzK-21000. Despite the large uncertainties, the MBB analysis indicates that the $\beta_{\rm eff}$ values of two galaxies are different at a $\sim2\sigma$ significance level. This difference in $\beta_{\rm eff}$ explains the similar effective $T_{\rm d}$ derived for the two sources despite the fact that the SED of GN20 peaks at $\sim90\,\mu$m while that of BzK-21000 peaks at $\sim100\,\mu$m. The best-fitting DL07 models suggest that BzK-21000 has properties similar to those found by Draine et al. (2007) for local spirals, with a larger fraction of dust in PAHs ($q_{\rm PAH}=3.9\%$) and a less intense radiation field ($U_{min}=8$) compared to that in GN20 ($q_{\rm PAH}=1.12\%$, $U_{min}=25$). Similar results are obtained by Elbaz et al. (2011).

Several studies have shown that there is a correlation between \gd ~and the enrichment of the ISM of a galaxy. 
In Fig. 3 (left) we plot \gd ~vs metallicity for a local sample studied by Leroy et al. (2011). For consistency we have computed all 
metallicities on Pettini \& Pagel 2004 (PP04) scale. This tight correlation (dispersion of $\sim0.15$ dex) 
can be used as a tool to constrain \aco. In particular, if we know the metallicity of a galaxy and
have measured its  \md, then we can estimate its \mgas ~and subsequently the  \aco ~value of the source, if \lco ~is known. In what
follows we will attempt to apply this approach for the two galaxies in our sample, under the assumption that the observed
\gd$-$metallicity relation for local galaxies holds at high redshift. 

Having derived estimates for the \md ~values of our sources, 
we need information on their metallicities, for which we have to rely on
indirect indicators. 
One of these is the $M_\ast$ $-$ metallicity relation of Erb et al. (2006), based on which BzK-21000 
has a slightly sub-solar metallicity, $Z=8.65$. A similar derivation would follow using the fundamental metallicity relation (FMR) of Mannucci et al. (2010) that relates the SFR and the stellar mass to metallicity.

For GN20, the situation is somewhat more complicated, as the Erb et al. (2006) relation is not sufficiently sampled
at the high mass end ($>10^{11}$ \msol). Using the FMR, and accounting for the evolution observed for galaxies at $z>$ 2.5 (Mannucci et al. 2010), we derive a metallicity of $Z=8.8$. Another metallicity estimate can be obtained by assuming that the huge SFR of GN20 could stem from the final burst of star formation, triggered by a major merger that will eventually transform the galaxy into  a massive elliptical. Once star formation ceases, the mass and metallicity of the resulting galaxy will not change further, and one might therefore apply the mass-metallicity relation for present-day elliptical galaxies. In this scenario, the metallicity of GN20 could range between $Z=8.8$ and $9.2$, although a moderately super-solar metallicity 
is more probable. For our purposes we will adopt a metallicity of $8.65\pm0.2$ for BzK-21000 and a whole range of $Z=8.8-9.2$ for GN20.

The fit to the local \gd$-Z$ relation (Fig. 3 left), indicates a value of \gd ~$\sim104$ for BzK-21000 and $\sim75$ (35) for GN20 assuming a
metallicity of $Z=8.8 (9.2)$. Santini et al. (2010) report similar \gd ~$\sim50$ for SMGs. Based on these ratios we first calculate \mgas ~for each galaxy and find $8.9 \times 10^{10}$ \msol ~and $1.5 \times 10^{11}$ \msol ~($7.0 \times 10^{10}$ \msol) for BzK-21000 and GN20 respectively. Then
based on the relation \mgas ~= \aco $\times$ \lco ~we estimate the corresponding \aco ~factors and find  \aco ~= 4.1$^{+3.3}_{-2.7}$ for BzK-21000 and in the case of GN20 \aco ~= 0.9$^{+0.4}_{-0.5}$ for $Z=8.8$ and 0.4$^{+0.2}_{-0.2}$ for $Z=9.2$. The derived values, along with the estimates of Leroy et al. (2011) for a local sample, are shown in Fig. 3 (middle). We also overplot a sample of local ULIRGs from Solomon et al. (1997), for which we were able to compute their metallicities on the PP04 scale. 
The quoted uncertainties account both for the dispersion of the \gd$-Z$ relation and the uncertainties in \gd ~and \md.  

The derived \aco ~values  agree with previous independent estimates. In particular  Daddi et al. (2010),
based on the dynamical masses of a sample of $z\sim1.5-2.0$ BzK galaxies (including BzK-21000), argued for an
average conversion factor of \aco ~= 3.6 for high$-z$ star forming disks and reported 
a value of \mgas ~= $(8.1\pm 1.4) \times 10^{10}$ \msol ~for BzK-21000 (see also Aravena et al. 2010). Furthermore, 
Carilli et al. (2010), based on the CO[1$-$0] transition line, estimated the gas
mass of GN20 to be \mgas ~= 1.3 $\times$ 10$^{11}$ $\times$ (\aco/0.8) \msol, putting a coarse upper limit on the
conversion factor of \aco ~$\sim0.8$ from dynamical constraints. This indicates that our \aco ~$<$ 1.0 estimate for
GN20 is reasonable. Inverting this line of reasoning,  placing the sources 
on the \gd ~$-$ metallicity plane of Fig. 3 (left) based on
the \mgas ~estimates from the literature, indicates that BzK-21000 is very close to the relation defined by a linear
regression fit to the local sample. Similarly, GN20  appears to
broadly follow the local trend. Another way to show that the \gd$- Z$ relation holds for high-z galaxies, is to 
plot direct observables, without any assumptions for \aco. Indeed, in Fig. 3 (right), we 
plot \lco/\md ~vs metallicity for our sample and find again that both GN20 and BzK21000 follow the local trend. Finally, recent results from Genzel et al. (2011) seem to verify our assumption.

Despite the substantial uncertainties, the agreement with independent \aco ~estimates is reassuring. 
In the local Universe there is observational and theoretical evidence that \aco ~in starburst galaxies 
is significantly smaller than that in the Milky Way disk (Downes \& Solomon 1998; Scoville et al. 1997). 
The derived conversion factor for GN20 is consistent with that of
local ULIRGs or even lower, while for BzK-21000 the value is considerably higher and close to that of local spirals.
We stress that without the addition of the mm data, the large
uncertainties in \md ~would not allow us to derive any meaningful conclusions. Furthermore, although the sources appear to have
comparable \lir/\lco ~$\sim100$ \lsol ~(K Km s$^{-1}$ pc$^{2}$)$^{-1}$, their SFEs, are considerably different. In particular, for BzK-21000 we derive a SFE $\sim25$ while for
GN20 SFE $\sim100-200$ (depending on the assumed metallicity).The two sources also have
comparable $M_{\ast}$/\md ~$\sim100$. This is also the case for local ULIRGs and normal SDSS galaxies (da Cunha et
al. 2010), supporting the idea that the property that distinguishes starbursts from normal star-forming galaxies is their
enhanced SFR (at fixed $M_{\rm gas}$, \md ~and $M_{\ast}$). 

We conclude by noting that our results, although limited to two sources, are in line with previous claims that the star-formation mode of BzK-21000 and other high$-z$ galaxies in the main sequence of the SFR$-$M$_{\ast}$ plane is different than that of most SMGs, and more similar to that of local-disks, despite their very large infrared luminosities and star formation rates. Additionally, we confirm the validity of the widely adopted
ULIRG-like \aco ~factor for SMGs (e.g., Tacconi et al. 2008). We have also demonstrated that the combination of \h ~with ground based millimetre data provides a powerful tool to investigate the dust and gas properties of high$-z$ galaxies. A larger sample is needed to extend this investigation and draw statistically robust conclusions. 

\begin{acknowledgements}
We thank Alvio Renzini for useful discussions. GEM acknowledges support from Oxford University.  
ED acknowledges funding support from
ERC-StG grant UPGAL 240039 and ANR-08-JCJC-
0008. DR acknowledges support from NASA through a Spitzer Space Telescope grant. This
work is based on observations made with the Herschel 
Space Observatory, a European Space Agency Cornerstone
Mission with significant participation by NASA.

\end{acknowledgements}

\begin{table*}
\centering
\caption{Summary of \h ~and (sub)mm data.}
\label{catalog}
\renewcommand{\footnoterule}{}  
\begin{tabular}{lcc}
\hline 
 $\lambda$ $\mu$m& GN20&  BzK-21000 \\
&[{\rm mJy}]&[{\rm mJy}]\\
\hline
100 & 0.7$\pm$0.4~$^a$& 8.1$\pm$0.6~$^a$\\
160&5.4$\pm$1.0~$^a$ &15.1$\pm$1.4~$^a$\\
250&18.6$\pm$2.7~$^a$&24.4$\pm$1.5~$^a$\\
350&41.3$\pm$5.2~$^a$&20.1$\pm$4.7~$^a$\\ 
500&39.7$\pm$6.1~$^a$&11.6$\pm$7.4~$^a$\\
850&20.3$\pm$2.0~$^b$&-\\
1100&10.7$\pm$1.0~$^c$&-\\
1300&-&0.87$\pm$0.32~$^d$\\
2200&0.90$\pm$0.15$^e$&0.32$\pm$0.15~$^f$\\
3300&0.33$\pm$0.06$^g$&0.04$\pm$0.06$^g$\\
6600&-0.01$\pm$0.018$^h$&-\\
\hline
\end{tabular}
\begin{flushleft}
{References:
$^a$ this paper; 
$^b$ Pope et al. (2006);
$^c$ Perera et al. 2008;
$^d$ Dannerbauer et al. (2011), in preparation;
$^e$ Dannerbauer et al. (2009);
$^f$ Carilli et al. (2010);
$^g$ Daddi et al. (2009);
$^h$ Carilli et al. (2011)
}
\end{flushleft}
\end{table*}

\begin{table*}
\centering
\caption{Summary of derived properties of GN20 and BzK-21000 with DL07 and MBB models.
}
\label{catalog}
\renewcommand{\footnoterule}{}  
\begin{tabular}{lccccccc}
\hline 
Object&\lir &$\chi^{2}_\nu$ (DL07)&\md (DL07)& $\chi^{2}_\nu$ (MBB)&\md (MBB)&\td &$\beta$  \\
&[\lsol] &&[10$^{8}$\msol]&&[10$^{8}$\msol]& [K]&\\
GN20&(1.9$\pm$0.4) $\times$ 10$^{13}$&2.14&$21.0^{+0.7}_{-0.6}$&1.54&$15.0^{+0.6}_{-0.5}$&32.6$\pm$2.2&2.1$\pm$0.2\\
BzK21&(2.1$\pm$0.3) $\times$ 10$^{12}$&1.21&$8.6^{+0.6}_{-0.9}$&0.87&$7.6^{+1.2}_{-1.3}$&33.8$\pm$2.1&1.4$\pm$0.2\\
\hline
\end{tabular}
\begin{flushleft}
\end{flushleft}
\end{table*}

\begin{figure*}
\centering
\includegraphics[scale=0.6]{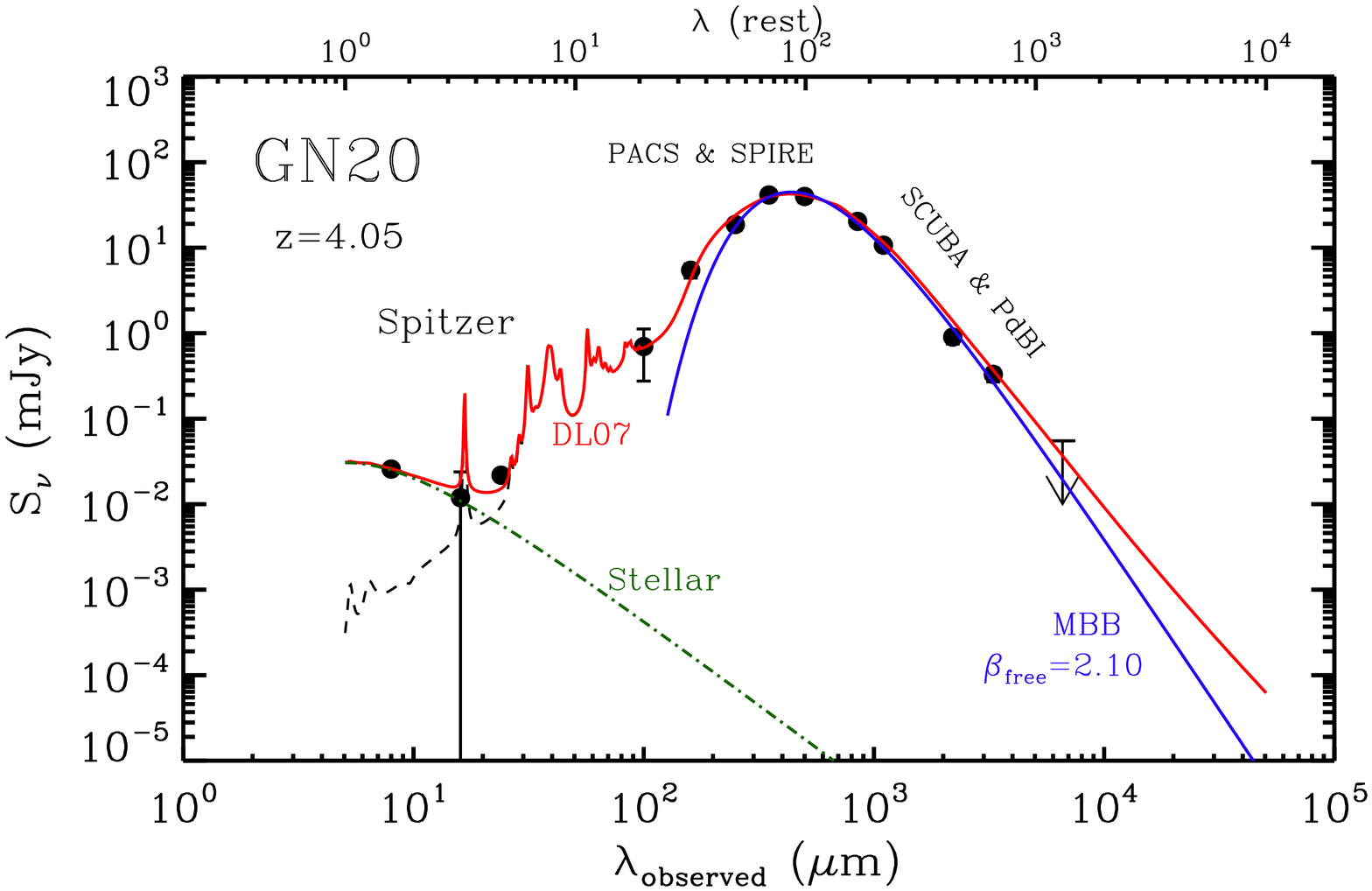}\\
\includegraphics[scale=0.6]{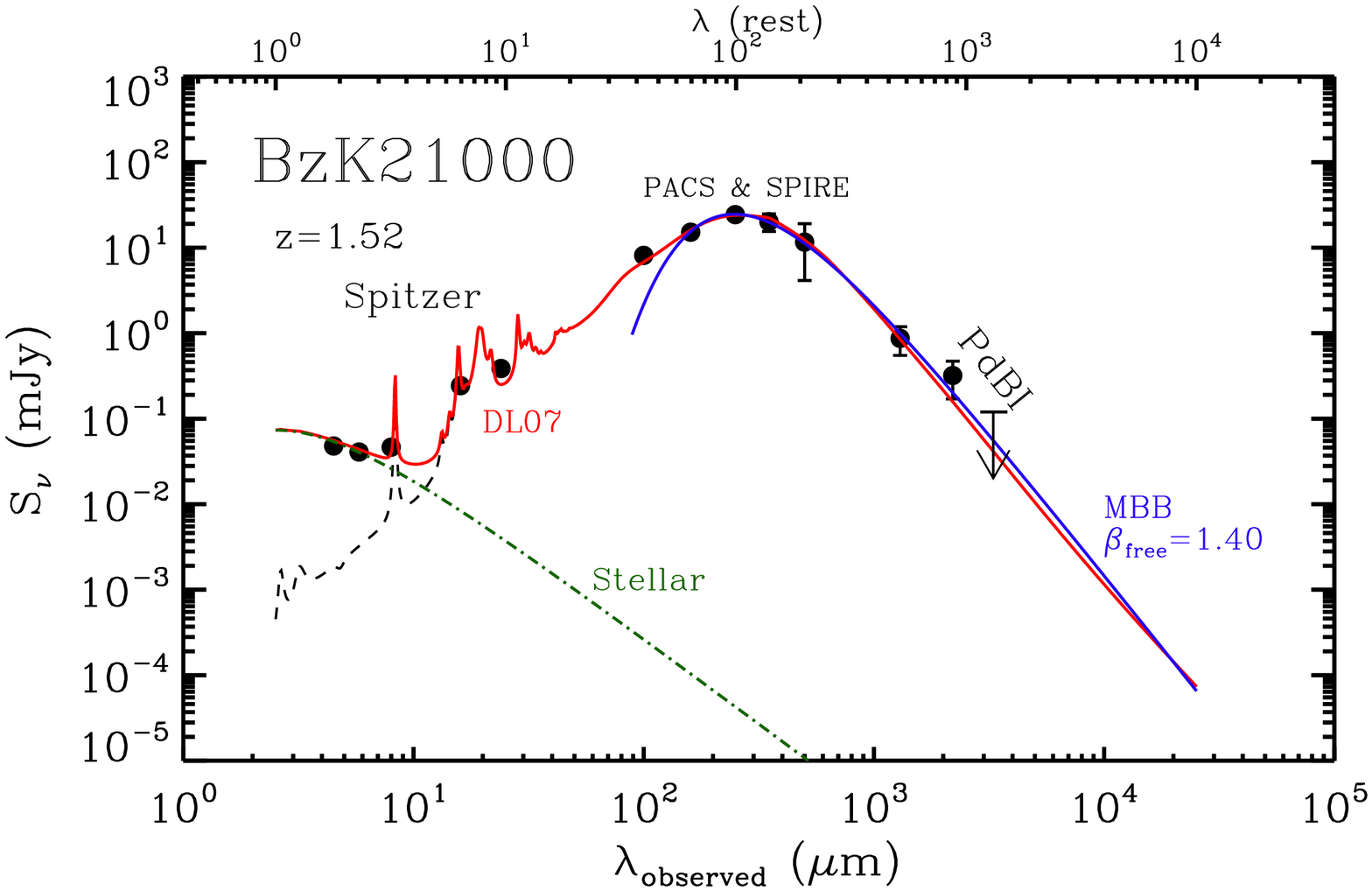}
\caption{Observed SED of GN20 (top) and BzK-21000 (bottom) overlaid with the best fit Draine \& Li (2007) (DL07) models (red) and the best fit  single temperature modified black body (blue). The black dashed line is the DL07 model without the stellar component, that is depicted with an green dotted-dashed line. The black arrows indicate the $3\,\sigma$ upper limit at 6.6$\,$mm and 3.3$\,$mm for GN20 and BzK-21000 respectively.}
\label{fig:sub} %
\end{figure*}

\begin{figure*}
\centering
\includegraphics[scale=0.34]{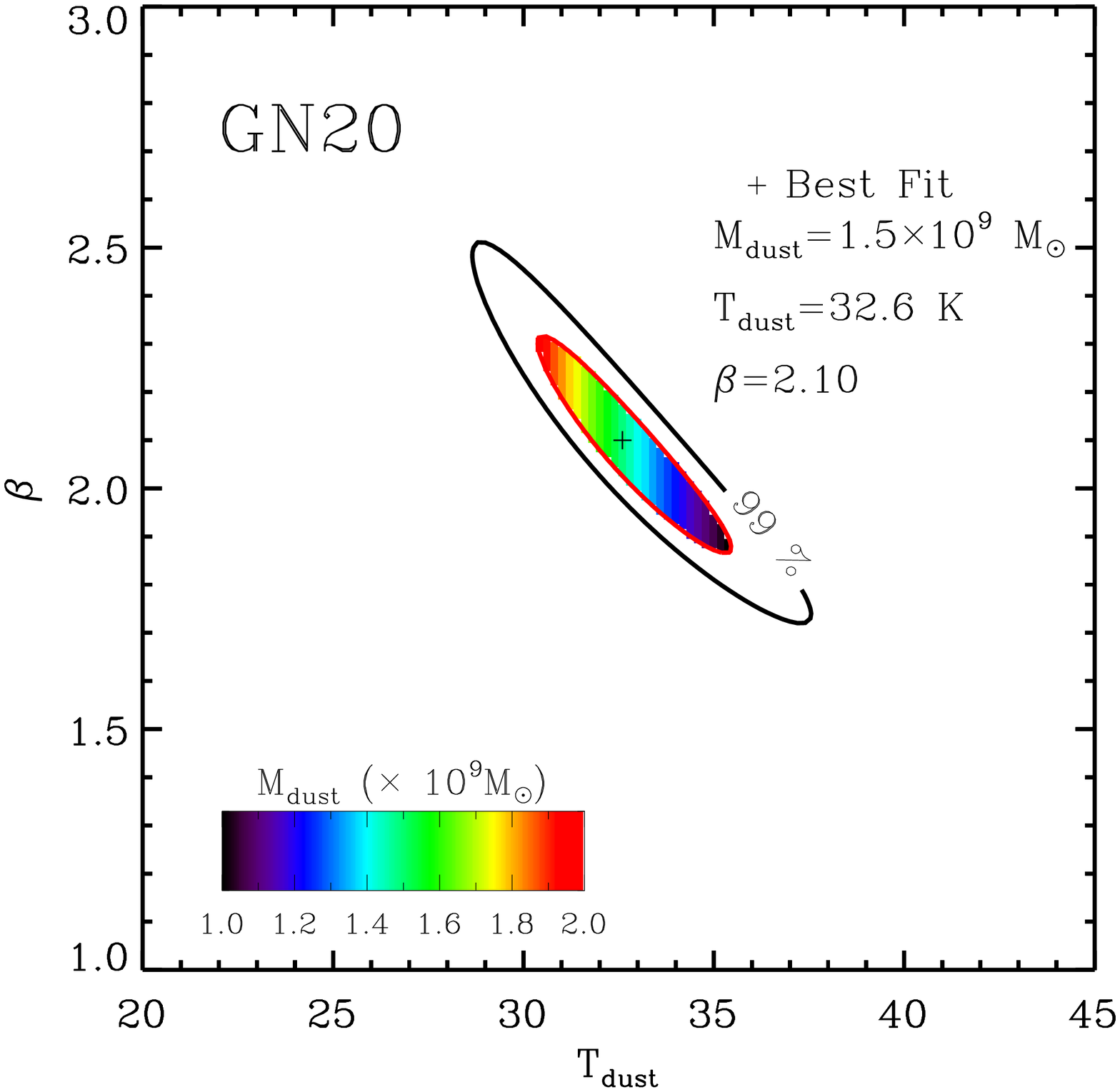}
\includegraphics[scale=0.34]{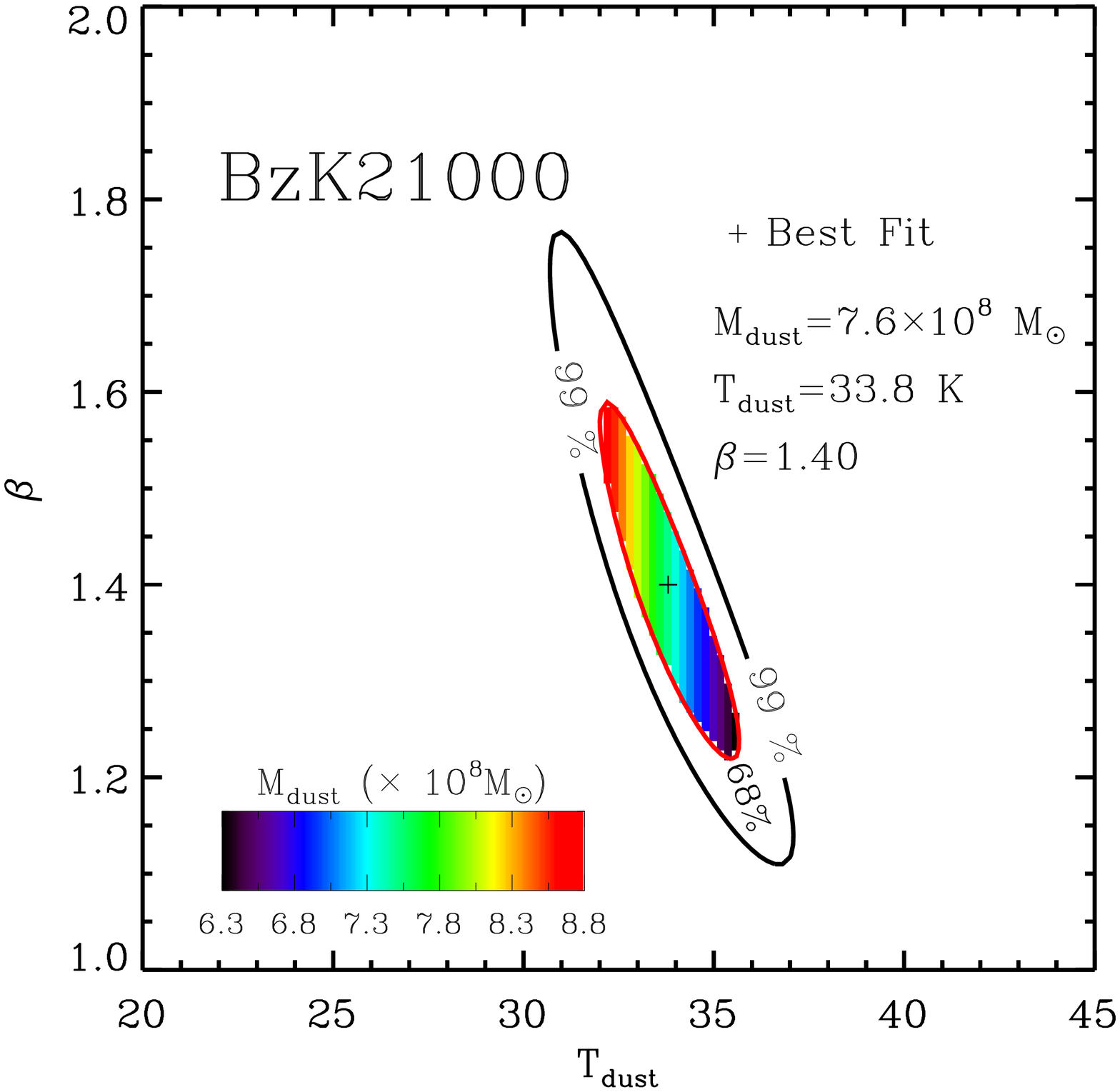}
\caption{Uncertainties on the parameters derived by MBB models for the case of GN20 (left) and BzK-21000 (right). The plot shows the equivalent 68\% and 99\% confidence intervals on $\beta_{\rm eff}$ plotted against \td, as derived by MC simulations. The region enclosed by the 68\% confidence level contour is colour coded based on the \md ~that corresponds to each set of $\beta_{\rm eff}$ and \td. The best fit value is denoted with a solid black cross (see also footnote 3).}
\label{fig:sub} %
\end{figure*}

\begin{figure*}
\centering
\includegraphics[scale=0.25]{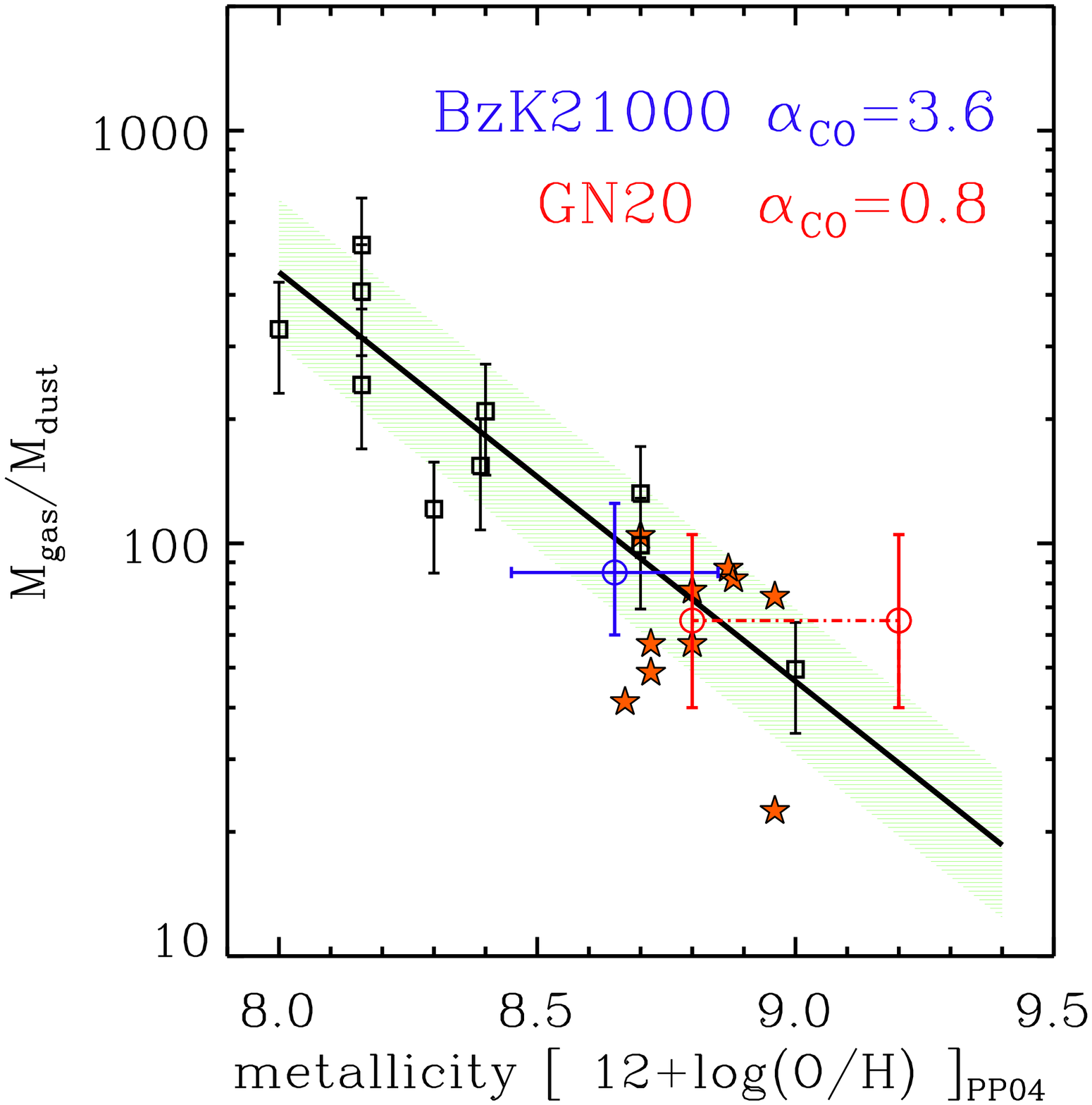}
\includegraphics[scale=0.25]{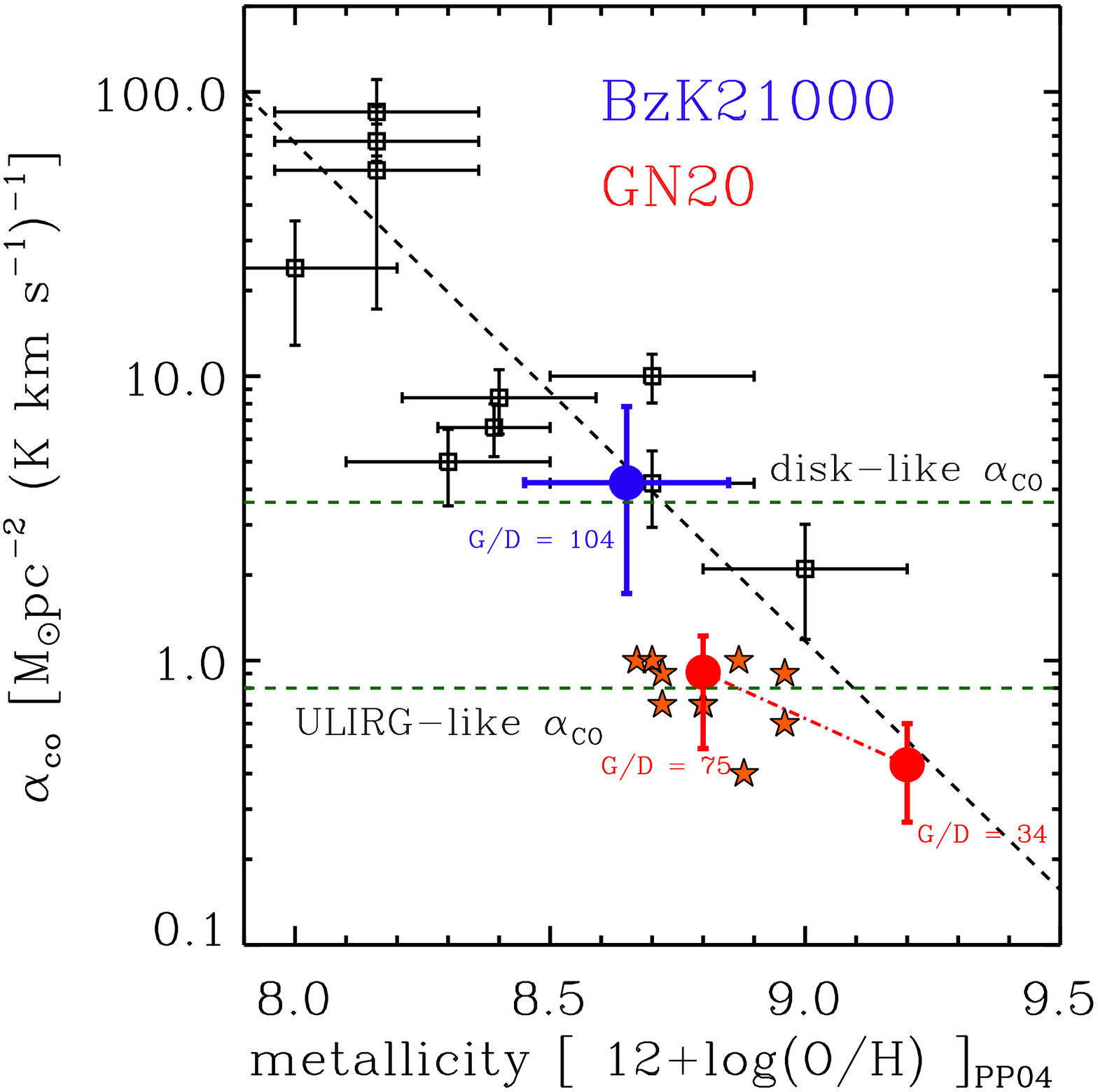}
\includegraphics[scale=0.25]{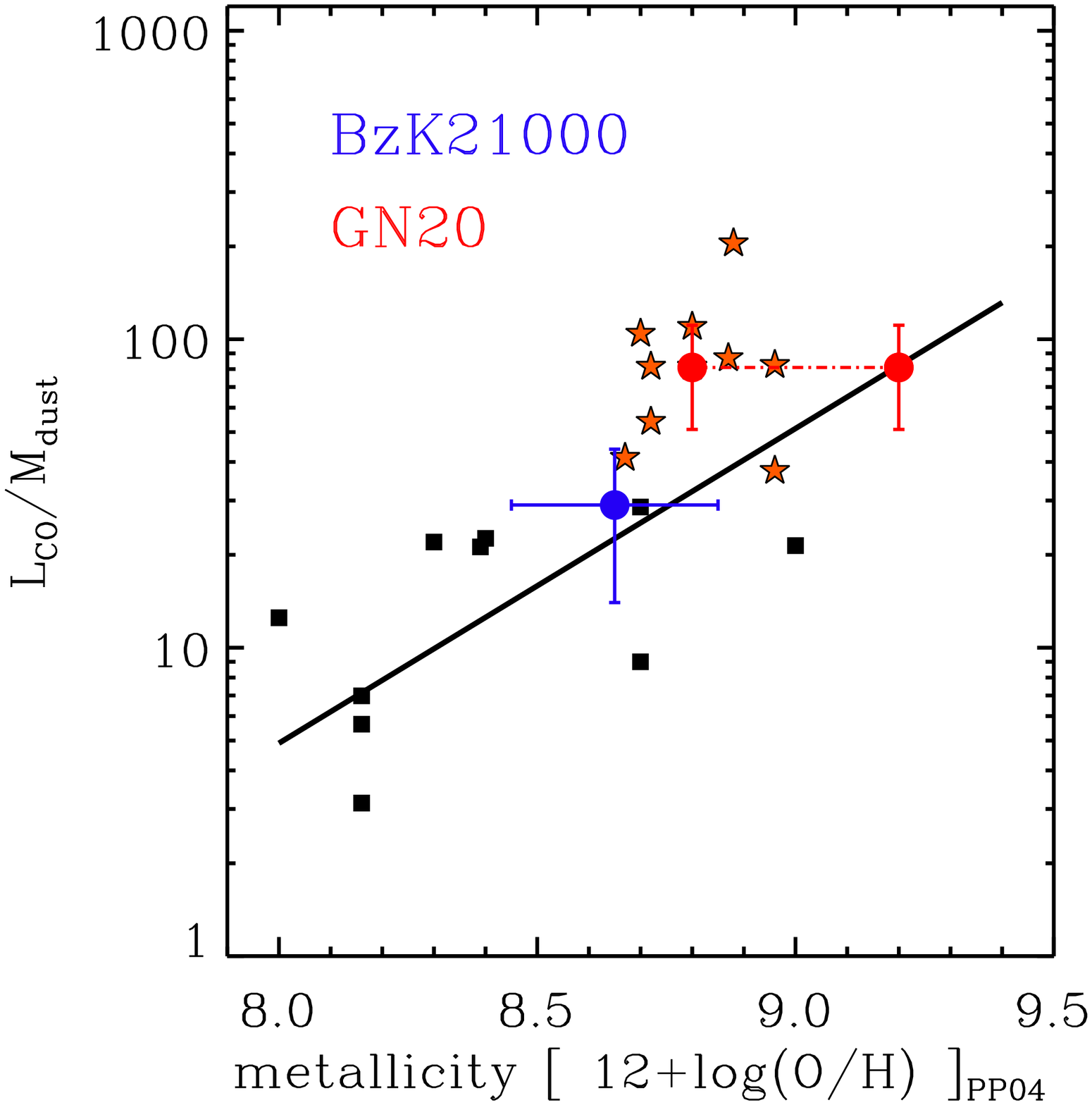}
\caption{Left: $M_{\rm gas}/M_{\rm d}$ vs metallicity for a sample of galaxies in the Local Group by Leroy et al. (2011) (black squares)  
and local ULIRGs by Solomon et al. (1997) (orange stars). The solid black line the best linear regression fit to Leroy's sample and the green shadowed area depicts the dispersion of the correlation. Blue and red circles indicate the position of BzK-21000 and GN20 respectively, based on
previously published \mgas ~values that correspond to \aco ~$=3.6$ for BzK-21000 (Daddi et al. 2010) and \aco ~$=0.8$ for GN20 (Daddi et al. 2010a, Carilli et al. 2010). GN20 is placed at $Z=8.8$ and $Z=9.2$ which are the lower and upper limits for its metallicity. All metallicities are calculated on PP04 scale. Middle: Constraints on \aco ~based on the local \gd$-Z$ relation shown in the left panel. Colors and symbols are the same as in the left panel. Right: \lco/\md ~ vs metallicity for the same sample of galaxies.}
\label{fig:sub} %
\end{figure*}

\clearpage

\end{document}